\documentclass{elsart}
\usepackage{amssymb}
\usepackage{epsfig}
\usepackage{wrapfig}
\usepackage{subeqn}
\newcommand{\Psfig}[2]{\epsfxsize=#1\epsfbox{#2}}
\newcommand{\Slash}[1]{\!\not{\!{#1}}}
\newcommand{\delsl}{\!\!\not{\!\partial}}
\newcommand{\barpsi}{\overline{\!\psi}}
\newcommand{\Ss}[1]{{\scriptscriptstyle #1}}
\newcommand{\rhoB}{\rho_\Ss{B}}
\newcommand{\rhoC}{\rho_\Ss{C}}
\newcommand{\rhoL}{\rho_\Ss{L}}
\newcommand{\muB}{\mu_\Ss{B}}
\newcommand{\muC}{\mu_\Ss{C}}
\newcommand{\muL}{\mu_\Ss{L}}

\newcommand{\MeV}{\mbox{MeV}}
\newcommand{\fmcube}{\hbox{fm}^{-3}}
\newcommand{\comment}[1]{}

\begin{document}
\begin{frontmatter}
\title{
Liquid-Gas Phase Transition \\
of Supernova Matter\\
and Its Relation to Nucleosynthesis
}

\author[Hokkaido]{C. Ishizuka}, 
\author[Hokkaido]{A. Ohnishi} 
and 
\author[Numazu]{K. Sumiyoshi}
\address[Hokkaido]{%
	Division of Physics, Graduate School of Science, 
	Hokkaido University, Sapporo 060-0810, Japan}
\address[Numazu]{%
	Numazu College of Technology,
	Numazu 410-8501, Japan}
\begin{abstract}
We investigate the liquid-gas phase transition of dense matter in supernova 
explosion by the relativistic mean field approach
and fragment based statistical model.
The boiling temperature is found
to be high
($T_{boil} \geq 0.7\ \MeV$ for $\rhoB \geq 10^{-7}\ \fmcube$), 
and adiabatic paths are shown to go across the 
boundary of coexisting region
even with high entropy.
This suggests that materials experienced phase transition
can be ejected to outside.
We calculated fragment mass and isotope distribution
around the boiling point.
We found that heavy elements at
the iron,
the first, second, and third peaks of r-process are abundantly formed
at $\rhoB = 10^{-7}, 10^{-5}, 10^{-3}$ and $10^{-2}\ \fmcube$, respectively.
\end{abstract}

\end{frontmatter}

\section{Introduction}

It is generally believed that there exist several phases in nuclear matter.
Among the phase transitions between these phases,
the nuclear liquid-gas phase transition has been extensively studied
in these three
decades~\cite{HI-Exp}.
It takes place in relatively cold ($T_{boil}$ = (5-8) MeV) and
less dense ($\rhoB\sim\rho_0/3$) nuclear matter,
and it causes multifragmentation
in heavy-ion collisions.
When the expanding nuclear matter cools down and goes across the 
boundary of coexisting region,
it becomes unstable against small fluctuations of density
or $np$ asymmetry, then various fragments are abundantly formed 
almost simultaneously.
Especially at around the critical point, 
fragment distribution is expected to follow
the power law~\cite{Fisher}, $Y_f \propto A^{-\tau}$,
which is one of the characteristic features of critical phenomena.
Recent theoretical model
studies~\cite{Dyn-HI,NSE-HI,Hirata2002,Fai-Randrup,Bauer}
have shown that it is very difficult to describe
this fragment distribution in a picture of 
sequential binary decays of one big compound nucleus,
which has been successfully applied to the decay of nuclei at low excitation.
This finding suggests that it is necessary to consider 
{\em statistical ensemble of various fragment configurations}
rather than one dominant configuration 
in describing fragment formation at around the boundary of
coexisting region.
%

In the universe, 
the temperature and density of this liquid-gas phase transition
would be probed during supernova explosion.
In the collapse and bounce stages of supernova explosion,
the density and temperature are high enough to keep statistical 
equilibrium~\cite{Bet90,Suz90}.
At baryon densities of $\rhoB \ge 10^{-5}\ \fmcube $,
since the density is too high for neutrinos to escape,
neutrinos are trapped in dense matter.
This leads to an approximate conservation of lepton fraction $Y_L = L/B$
and entropy per baryon $S/B$,
where $L$ and $B$ denote the lepton and baryon number, respectively.
After the core bounce, supernova matter, 
which is composed of nucleons and leptons,
expands and cools down.
As the baryon density and the temperature decrease,
charged particle reactions become insufficient
and the chemical equilibrium ceases to hold,
namely the system freezes out at this point.
If the supernova matter goes across the boundary of coexisting region 
and the boiling point $T_{boil}$ of the liquid-gas phase transition
is higher than the freeze-out temperature $T_{fo}$,
this matter will dissolve into fragments
and form various nuclei in a critical manner.
It further keeps equilibrium and expands to the freeze-out point.
The statistical distribution of fragments at freeze-out would
provide the initial condition for following nucleosynthesis
such as the r-process. (See following references on 
r-process~\cite{Burbidge,Meyer} and references therein.)

The importance of the nuclear liquid-gas phase transition in 
supernova explosion was already noticed and extensively
studied before~\cite{Lattimer}.
However, there are two more points which we should consider further.
First, the main interest in the previous works
was limited to the modification of the equation of state (EoS).
The nuclear distribution as an initial condition for the r-process
was not studied well.
Secondly, in constructing the EoS of supernova matter,
the mean field treatment was applied in which one assumed 
one kind of large nucleus surrounded by nucleon and alpha 
gas~\cite{TM1-table,LS91}.
At temperature much above or below the boiling point,
fragment mass distribution is narrow and 
the one species approximation works well.
However, since fluctuation dominates at around the boiling point,
it is  necessary to take account of fragment mass and isotope distribution.
This distribution of fragments can modify the 
following r-process nucleosynthesis
provided that the freeze-out point is not far from the boiling point.

In this work, 
we study nuclear fragment formation
through the nuclear liquid-gas phase transition
during supernova explosion.
This process may lead to the production of medium mass nuclei
as seed elements and serve as a pre-process of the r-process.
We call this process as LG process~\cite{IOS2001-YKIS01b}.
%
%
In order to pursue this possibility quantitatively, 
it is necessary to determine the liquid-gas
coexisting region.
We find that the liquid-gas coexisting region extends down to very low density
keeping the boiling point around $T_{boil} \sim 1$ MeV
in a two-phase coexistence treatment of EoS 
with the Relativistic Mean Field (RMF)
model~\cite{TM1-table,Sero-Walecka,TM1,Muller,RMF-Other}.
In supernova explosion, it can happen that
material with $S/B \geq 10$ is ejected to outside~\cite{SumiyoshiNext}.
Adiabatic paths of ejecta are found to go through
the calculated liquid-gas coexisting region even with high entropy.
%
%
Having this finding of the passage through the coexisting region,
we investigate the fragment distribution at around the boiling point
in a statistical models of
fragments~\cite{NSE-HI,Hirata2002,Fai-Randrup,Bauer},
referred to as the Nuclear Statistical Equilibrium (NSE) in astrophysics. 
We show that heavy elements around the first, second,
and third peaks of r-process are abundantly formed
at $\rhoB = 10^{-5}, 10^{-3}$ and $10^{-2}\ \fmcube$, respectively, 
with temperatures around and just below $T_{boil}$ in NSE.
Furthermore, the isotope distribution of these elements
are also well described in this model.
We find that it is important to take account of 
the Coulomb energy reduction from the screening by electrons
in supernova matter.
Although nuclei formed at high densities $\rhoB \sim 10^{-2}\ \fmcube $
having very small entropy at around $T_{boil}$
is not likely ejected to outside,
heavy elements up to the r-process third peak
are already formed statistically at these densities.

This paper is organized as follows.
We describe the treatment of
two-phase coexistence 
with RMF model in Sec. 2.
The liquid-gas coexistence is shown in the ($\rhoB,T$) diagram.
We demonstrate that it would be possible for a part of ejecta
in supernova explosion to experience the liquid-gas coexisting region.
The effects of liquid-gas coexistence on EoS,
proton fraction, and adiabatic path are also studied.
In Sec. 3, 
we describe 
the nuclear statistical model of fragments at equilibrium (NSE)
to study the production of elements.
We take into account the Coulomb energy modification from electron screening.
We evaluate the fragment distribution at around 
the boiling point and in the coexisting region
within this statistical model.
We found that fragments are formed abundantly 
even at very low densities if the temperature is around the boiling point.
We compare the calculated mass and isotope distributions of fragments
with the solar abundance~\cite{Abundance}.
In Sec. 4,
we discuss the possibility of
the ejection of nuclei synthesized 
in the coexisting region 
referring to a hydrodynamical calculation of supernova 
explosion~\cite{SumiyoshiNext}.
We summarize our work in Sec. 5.


\vfill\break

\section{Relativistic Mean Field Approach}

\subsection{RMF Lagrangian and parameter set}

Relativistic Mean Field (RMF) approach has been developed
as an effective theory to describe the nuclear matter saturation
in a simple way~\cite{Sero-Walecka}.
Having improvements to include meson self-coupling terms,
it describes well the binding energies of neutron-rich unstable nuclei
in addition to nuclear matter and stable nuclei~\cite{TM1}.

In this work, we adopt an RMF parameter set TM1~\cite{TM1}.
It has been demonstrated that this parameter set TM1 
can reproduce nuclear properties
including proton- and neutron-rich unstable nuclei.
In addition, the EoS table with TM1 has been successfully applied
to neutron stars and supernova 
explosions~\cite{TM1-table,Sum95b,Sum00,Sum95c}.
Therefore, it is expected to be reliable also in describing two-phase
coexistence in supernova matter,
which contains asymmetric nuclear matter having
proton-to-neutron ratio varying in a wide range.

The Lagrangian contains three meson fields; 
scalar-isoscalar $\sigma$ meson,
vector-isoscalar $\omega$ meson,
and vector-isovector $\rho$ meson.
In this work, we limit the constituent particles as
nucleons, electrons, electron-neutrinos, their anti-particles
and photons.
The explicit form of the Lagrangian including leptons
is given as follows.
\begin{eqnarray}
\mathcal{L}
	&=& \barpsi_N
	    \left(i\delsl-M
		-g_{\sigma}\sigma
		-g_{\omega}\Slash{\omega}
		-g_{\rho}\tau^{a}\Slash{\rho}^{a}
	    \right)\psi_N
\nonumber \\
	&+& \frac{1}{2}\partial^{\mu}\sigma\partial_{\mu}\sigma
		- \frac{1}{2}m_{\sigma}^2\sigma^2
	- \frac{1}{3}g_2\sigma^3 - \frac{1}{4}g_3\sigma^4
\nonumber \\
	&-& \frac{1}{4}W^{\mu\nu}W_{\mu\nu}
		+ \frac{1}{2}m_{\omega}^2\omega^{\mu}\omega_{\mu}
	- \frac{1}{4}R^{a\mu\nu}R_{\mu\nu}^a
		+ \frac{1}{2}m_{\rho}^2\rho^{a\mu}\rho_{\mu}^a
	+ \frac{1}{4}c_3\left(\omega_{\mu}\omega^{\mu}\right)^2
\nonumber \\
	&+&\barpsi_e
	    \left(i\delsl - m_e\right)\psi_e
	+\barpsi_\nu
		i\delsl
	    \psi_\nu
	-\frac{1}{4}F_{\mu\nu}F^{\mu\nu}    
\ ,\nonumber \\
W_{\mu\nu}
&=& \partial_\mu \omega_\nu - \partial_\nu \omega_\mu\ ,\nonumber \\
R^a_{\mu\nu}
&=& \partial_\mu \rho^a_\nu - \partial_\nu \rho^a_\mu 
		+ g_\rho\epsilon^{abc}\rho^{b\mu}\rho^{c\nu}\ ,\nonumber \\
F_{\mu\nu}
&=& \partial_\mu A_\nu - \partial_\nu A_\mu\ .
\end{eqnarray}
In this section, photon contributions to pressure, energy and
entropies are included, but we have dropped the photon-couplings
to charged particles (See Sec. 3 on this point). 

In a mean field approximation, 
three meson fields are replaced to their expectation values.
Self-consistency condition 
for these values can be derived in a standard manner,
as $\partial P/\partial x = 0$
where $x$ represents the meson field expectation value
and $P$ denotes the pressure.

Supernova matter is composed of nucleons, electrons, neutrinos, their 
anti-particles and photons,
characterized by a fixed lepton fraction (lepton-to-baryon ratio) $Y_L$,
due to the neutrino trapping at high density.
As a result, there are three conserved quantities; 
baryon number $B$, total charge $C$, and lepton number $L$.
Then the chemical potentials of particles are represented
by the corresponding three chemical potentials,
\begin{equation}
	\mu_i = B_i \muB + C_i \muC + L_i \muL\ ,
\end{equation}
where $\mu_B, \mu_C, \mu_L$ are baryon, charge and lepton chemical potentials,
and $B_i, C_i, L_i$ are baryon, charge and lepton numbers 
of particle species $i$.
The chemical equilibrium conditions for given conserved densities $\rho_k$
are expressed as $\partial P/\partial \mu_k = \rho_k$.
We solve these conditions in a multi-dimensional Newton's method by iteration.


\subsection{Two-phase coexistence treatment in RMF}

We apply the mean field approximation
to the liquid and gas phases separately.
In this treatment, we implicitly assume that 
two coexisting (liquid, gas) phases are uniform and have infinite size.
We solve chemical and thermal equilibrium conditions
between the liquid and gas phases.


In order to make liquid and gas phases coexist at equilibrium,
we must apply the Gibbs conditions rather than the Maxwell construction,
since the number of conserved quantity (= 3) is larger than one.
The Gibbs conditions are given as follows,
\begin{equation}
\left(1 - \alpha\right)\rho_k^{Liq.} + \alpha\rho_k^{Gas} = \rho_k
\ ,\quad
\mu_k^{Liq.} = \mu_k^{Gas}
\ ,\quad
P^{Liq.} = P^{Gas}
\ ,
\end{equation}
where $k = B, C, L$.
The quantities labeled by $Liq.$ and $Gas$ are those of liquid and 
gas phases, respectively.
The gas volume fraction, $\alpha$, is a number between 
zero and unity.
We solve these conditions by using multi-dimensional Newton's method,
in which the dimension is five in the asymmetric nuclear matter
($k = B, C$) and seven in the supernova matter ($k = B, C, L$),
where the variables are $\rho_k^{Liq.}, \rho_k^{Gas}$ and $\alpha$.
See Appendix A for the numerical technique at very low densities.

%

\begin{figure}[htbp]
\centerline{~\Psfig{12.0cm}{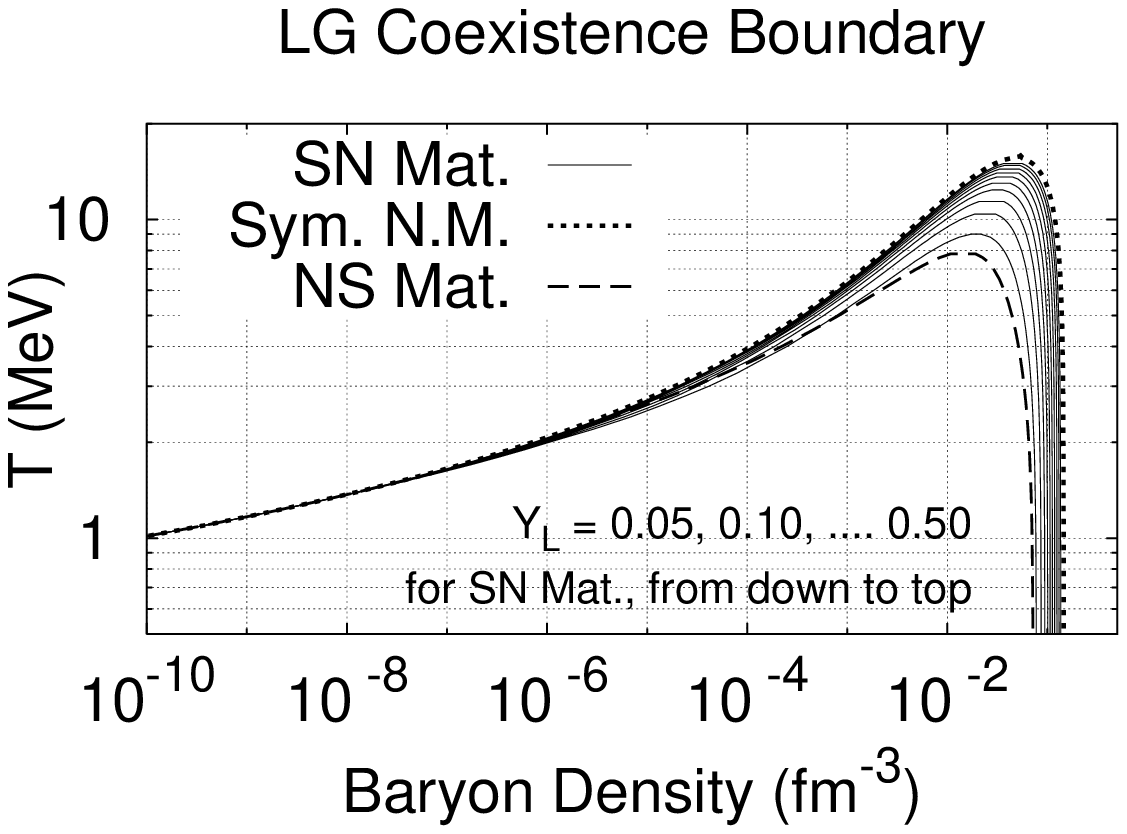}~}
\caption{
Boundary of liquid-gas coexisting region
in supernova matter (solid curves)
in comparison with symmetric nuclear matter ($Y_p = 0.5$, dotted line) 
and neutrino-less supernova matter (or finite temperature
neutron star matter, $\mu_L = 0$, dashed line).
}
\label{Fig:Tc}
\end{figure}

We show the liquid-gas coexisting region of supernova matter
in Fig.~\ref{Fig:Tc}.
See Appendix B for the method to determine this boundary.
The solid curves show the boundary of coexisting region,
and two phases coexist below the boundary.
We find that the 
critical temperature is very high $T_{c} \sim $14 MeV
for $Y_L = (0.3-0.4)$,
which is the ratio expected in actual supernova explosions.
These critical temperatures are much higher than those
in neutrino-less supernova matter (dashed line)
and comparable to those in the symmetric nuclear matter (dotted line).
The boiling points of supernova matter
remain to be $T_{boil} \sim$ 1 MeV even at very low densities
for all lepton fractions.
%

The large value of $T_{boil}$ 
is due to symmetrization of nuclear matter by leptons.
Since $Y_L$ is kept constant but $Y_p$ is not fixed,
the supernova matter searches its minimum in 
free energy by changing $Y_p$.
%
%
In high density supernova matter,
the lepton fraction is shared by electrons and neutrinos
($2/3Y_L < Y_e < Y_L$),
then the net neutrino fraction and the lepton chemical potential
become positive ($\muL = \mu_\nu > 0$).
Compared to neutrino-less supernova matter ($\muL = 0$),
this positive $\muL$ in supernova matter helps to enhance $Y_e (= Y_p)$
($\mu_e = -\muC + \muL$), and to symmetrize nuclear matter.
%
%
At very low baryon densities,
electrons become non-degenerate, i.e.
electron and anti-electron densities are much higher than
the net electron densities because of small electron mass,
and this also applies to neutrinos.
Hence, electron and neutrino chemical potentials become small,
and the charge chemical potential also becomes small.
This causes nuclear matter symmetric.

\begin{figure}
\centerline{~\Psfig{10.0cm}{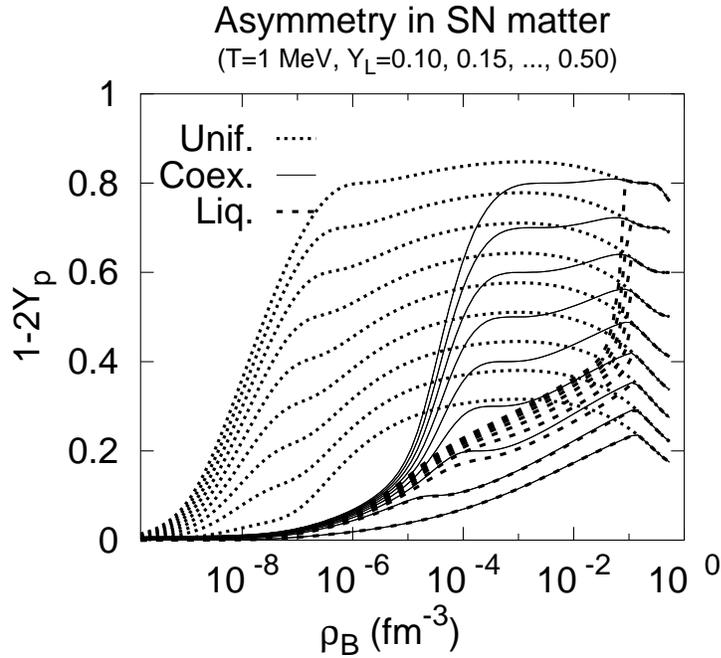}~}
\caption{
Asymmetry parameter $\delta = 1 - 2Y_p$, of supernova matter
as a function of baryon density at $T = 1$ MeV.
Dotted, solid and dashed curves show 
the asymmetry parameter
in uniform (homogeneous), 
two-phase coexisting, and the liquid part of coexisting matter,
respectively.
}
\label{Fig:Yp}
\end{figure}

In order to demonstrate this point, 
we show the density dependence of the asymmetry parameter
of supernova matter in Fig.~\ref{Fig:Yp}.
Dotted, solid and dashed curves show the asymmetry parameter
in uniform (homogeneous), 
two-phase coexisting, and the liquid part of coexisting matter,
respectively.
It is clear that, as the baryon density decreases,
asymmetry parameter decreases
and this tendency is stronger in the coexisting region.
%
%
In uniform matter, 
the asymmetry decreases rapidly
at around $\rhoB < 10^{-7}\ \fmcube$,
where leptons dominate pressure and energy density,
as shown later in Fig.~\ref{Fig:EoS}.
This is consistent with the above consideration on the lepton dominance.
%
%
On the other hand, the asymmetry parameter start to decrease at 
much higher density in coexisting matter.
This symmetrization is mainly due to the symmetry energy in nuclear matter.
Since the baryon density in the liquid part of matter is around $\rho_0$,
nucleons can gain symmetry energy by reducing  the asymmetry.
As shown by dashed lines, the asymmetry decreases quickly
in the liquid part in the coexisting region.

\begin{figure}
\centerline{~\Psfig{10cm}{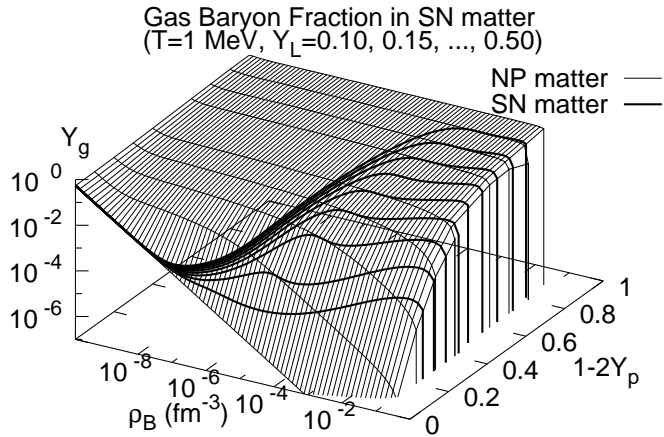}~}
\caption{
Gas fraction $Y_g$ in asymmetric nuclear matter (thin line surface)
and in supernova matter (thick lines)
as a function of the baryon density and the asymmetry, $\delta = 1 - 2 Y_p$,
at $T = 1$ MeV.
}\label{Fig:Yg-RMF}
\end{figure}

We turn our attention to one of the characteristic features of the 
liquid-gas coexisting region in supernova matter, which allows
many fragments to be formed even at very low densities. 
First we define a new quantity, gas fraction $Y_g$,
$Y_g$ = (baryons in gas phase)/(total baryons)
$\equiv \alpha \rhoB^{Gas}/\rhoB$
as a measure of bulk fragment yield.
Here, we can adopt the normal nuclear matter density $\rho_0$
for the baryon densities in the liquid phase.
After we substitute these baryon densities
$\rho_0$ and $\rhoB^{Gas}$ for the Gibbs condition, 
we can obtain the following equation,
\begin{equation} 
\alpha \simeq \frac{\rho_0 - \rhoB}{\rho_0 - \rhoB^{Gas}} \ .
\end{equation}
For example, all baryons are bound in nuclei at $Y_g$ = 0.
Smaller $Y_g$ means larger amount of fragments.
In Fig.~\ref{Fig:Yg-RMF},
we show the gas fraction in 
nuclear matter composed of neutrons and protons ($np$ matter)
and supernova matter
as a function of the baryon density and the asymmetry parameter,
calculated in the RMF model.
We fix $Y_p$ for calculations of
$np$ matter. 
The behavior of $Y_g$ in 
$np$ matter is smooth.
In symmetric nuclear matter ($Y_p = 0.5$), 
both of the liquid and gas phases are symmetric,
and the density in each phase is constant in the coexisting region.
Then the gas fraction can be expressed by liquid,
gas and the given average densities 
($\rhoB^{Liq.}\simeq\rho_0, \rhoB^{Gas}, \rhoB$) as
\begin{equation}
\label{Eq:YGsym}
Y_g \simeq \frac{\rhoB^{Gas}(\rho_0 - \rhoB)}{\rhoB(\rho_0 - \rhoB^{Gas})} \ .
\end{equation}
This is a monotonically decreasing function of the baryon density
and very small in the density range under consideration.
When the asymmetry increases, the liquid phase loses the symmetry energy
and nucleons are emitted to the gas phase,
while $Y_g$ is still a decreasing function of $\rhoB$.

The behavior of $Y_g$ in supernova matter is very different from
that in $np$ matter.
As shown by the thick lines in Fig.~\ref{Fig:Yg-RMF},
as the baryon density decreases, $Y_g$ first decreases then grows again.
This behavior is determined by the proton fraction (proton-to-baryon ratio)
$Y_p$.
In the medium density region ($\rhoB = (10^{-7}-10^{-2})\ \fmcube$),
matter becomes symmetric as $\rhoB$ decreases,
and the gas baryon fraction decreases
to the symmetric nuclear matter value in Eq.~(\ref{Eq:YGsym}).
Therefore, baryons favor liquid state than nucleon gas 
in the coexisting region.
After reaching symmetric matter, gas fraction increases again 
according to Eq.~(\ref{Eq:YGsym}).
%
It is also interesting to note that 
the proton fraction $Y_p$ of supernova matter shows
clear $Y_L$ dependence at $\rhoB > 10^{-5}\ \fmcube $,
while $Y_p$ is almost independent on $Y_L$
at $\rhoB < 10^{-5}\ \fmcube$.
This density roughly corresponds to the neutrino-sphere,
inside which neutrinos are trapped in supernova core.

\subsection{Supernova Matter Equation of State}

\begin{figure}
\centerline{~\Psfig{\textwidth}{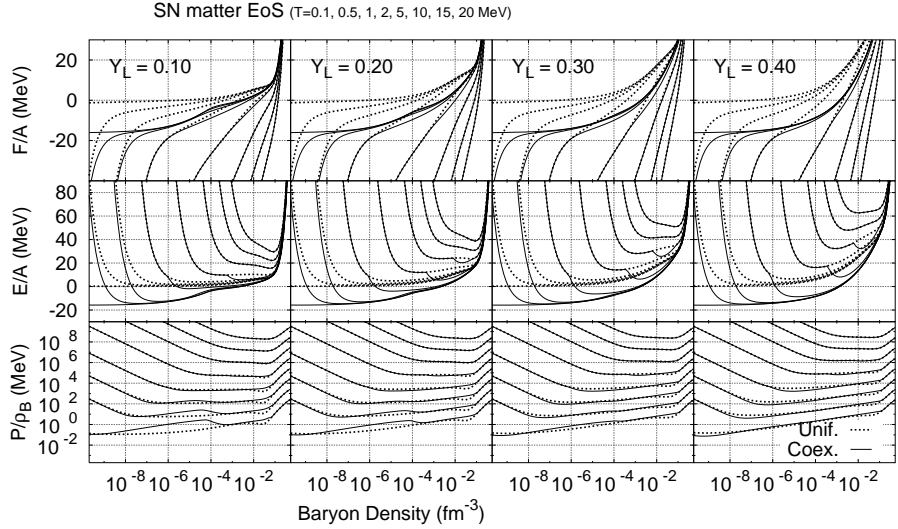}~}
\caption{Equation of state of supernova matter for lepton fractions
$Y_L =$ 0.1, 0.2, 0.3 and 0.4.
Temperatures are $T=$0.1, 0.5, 1, 2, 5, 10, 15 and 20 MeV,
from top to down (from down to top) for $F/A$ ($E/A$ and $P/\rhoB$).
Results for uniform supernova matter are shown by dotted curves,
and those with liquid-gas coexistence are shown by solid curves.
For energies per baryon $(E/A)$ and 
free energies per baryon $(F/A)$, nucleon mass is subtracted.
The pressures are increased by a factor of 10 from low to high temperatures
(from bottom to top).
}
\label{Fig:EoS}
\end{figure}

We show the EoS of supernova matter 
with (without) two-phase coexistence by solid (dotted) lines
in Fig.~\ref{Fig:EoS}.
In both of the cases, 
energy and pressure increase at high densities above $\rho_0$
because of the vector meson contributions.
At very low densities,
lepton contributions become dominant,
because lepton pressure and energy densities are finite 
even with $\rhoB = 0$ at finite temperatures.
In uniform matter, 
the nuclear pressure and energy become close to free-gas values
$P_N/\rhoB \to T, E_N/A \to 3T/2$.
Between these two extremes of density, 
we can find the effects of two-phase coexistence.

In two-phase coexistence,
by making liquid phase whose baryon density is around $\rho_0$,
binding energy is gained.
At low temperatures, since nuclear matter tends to be symmetric
and liquid phase is dominant in a wide range of densities,
liquid part of energy per baryon approaches toward $-16$ MeV.

At large lepton fractions ($Y_L = 0.3-0.4$),
the pressure behaves as in the case of Maxwell construction
for volume instability;
pressures in the coexisting region are lower (higher) at low (high) densities
than in uniform matter during the coexistence.
On the other hand, we can find double phase transition behavior
in lepton deficient matter ($Y_L = 0.1-0.2$);
while the pressure behaves as in the case of volume instability
at higher densities ($\rhoB \geq 10^{-4}\ \fmcube$,
there is another overtaking
in the density region of $\rhoB = 10^{-8} \sim 10^{-4}\ \fmcube$.
At these densities, 
the matter becomes unstable to the small fluctuations of proton fraction,
and liquid part of the matter becomes rapidly symmetric
as the density decreases.
Thus the overtaking of the pressure in the low-density region
may be suggesting the phase transition in the isospin degrees of freedom.


\subsection{Adiabatic Paths}

\begin{figure}
\centerline{~\Psfig{\textwidth}{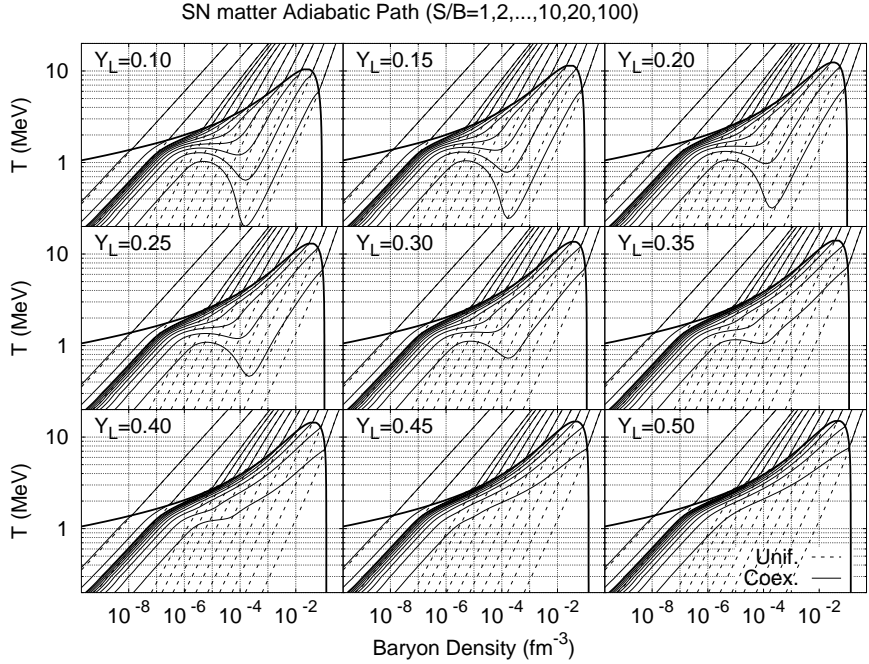}~}
\caption{
Adiabatic paths in supernova matter calculated in RMF.
The thin solid (dashed) lines show
the adiabatic paths of supernova matter
with (without) the liquid-gas coexistence.
Thick solid lines are the boundary of coexisting region.
The each adiabat corresponds to entropy per baryon $S/B=1, 2, ..., 10,
 20, 100$. The panels are in case of lepton fraction
$Y_L$ = 0.1, 0.15, ..., 0.5 from upper-left to lower-right, respectively.}
\label{Fig:ADP}
\end{figure}

After the core bounce,
some part of supernova matter expands almost adiabatically,
and high entropy ($S/B \geq 10$) part of the matter
would be ejected to outside~\cite{SumiyoshiNext}.
(See also Section 4.)
Therefore, it is important to examine supernova matter along
the adiabatic paths.
In Fig.~\ref{Fig:ADP}, 
we show the adiabatic paths in supernova matter
with lepton fraction $Y_L = 0.10, 0.15, \ldots 0.50$.
Entropy per baryon is taken to be
$S/B = 1, 2, \ldots 10, 20,$ and 100.
At very high entropies such as $S/B = 100$,
adiabatic paths are almost independent on $Y_L$ 
because of the lepton dominance.
At lower entropies,
we can clearly see nuclear and coexistence effects.
While adiabatic paths in uniform matter (dashed lines) evolve 
very smoothly,
those with two-phase coexistence (thin solid lines) bend
closer to the boundary of coexisting region.
This can be seen even at $S/B = 20$.
This bending comes from the latent heat and suggests that significant
amount of nuclei are formed around the boundary.


%
\section{Fragment Distribution in Supernova Matter}

Observations in the previous section tell us that it would be
possible for supernova matter to experience the liquid-gas 
phase transition before it is ejected to outside.
Therefore, it is interesting to study the composition of ejecta
experienced the phase coexistence.
However, we have assumed that the coexisting two phases are infinite.
It is also to be noted that
nuclear matter in the liquid phase is more symmetric than in the gas phase.
As a result, the Coulomb energy is expected to be large,
which we have neglected in the previous section,
and the infinite matter in the liquid phase 
will fragment into finite nuclei.
Although the effects of these nuclear formation 
on EoS may not be very large~\cite{Lattimer},
fragment distribution at freeze-out would serve as the initial condition
for the following r-process.
Therefore, in this section,
we evaluate the fragment yield in a fragment-based statistical model.

%
\subsection{Statistical Model of Fragments}

In order to describe distribution of finite nuclei,
we utilize a fragment-based statistical model.
This kind of statistical models have been widely used
in heavy-ion collision 
studies~\cite{NSE-HI,Hirata2002,Fai-Randrup,Bauer}
as well as in astrophysics,
as referred to the nuclear statistical 
equilibrium (NSE)~\cite{Meyer,Clayton}. 
In NSE,
we solve the statistical equilibrium condition among nucleons,
fragments and leptons in fragment-based grand canonical ensemble.
In this work, we have ignored relativistic corrections
and anti-particle contributions of fragments,
and fragments are assumed to follow the Boltzmann statistics,
while leptons are treated relativistically.
The ensemble averages can be generated from the grand potential,
\begin{eqnarray}
\label{eq:Omega}
\Omega 
	&=& - PV = -VT \sum_i \rho_f - P_\ell V - P_\gamma V
\ ,\\
\label{eq:rho}
\rho_f
	&=& \zeta_f(T) \left({M_f T \over 2\pi\hbar^2}\right)^{3/2}
			\exp\left({B_f + \mu_f \over T}\right)
\ ,\\
\label{eq:mu}
\mu_f &=& Z_f (\mu_p - m_N) + N_f (\mu_n - m_N)
\ ,
\end{eqnarray}
where $\rho_f$, $M_f$, $B_f$ and $\mu_f$ denote
the density, mass, binding energy and the chemical potential
of fragment $f$,
and $P_\ell$ and $P_\gamma$ are the lepton and photon pressures, 
respectively.
The intrinsic partition function, $\zeta_{f}(T)$, 
has been calculated by using the level density formula
for fragments with $A_f \ge 5 $~\cite{Fai-Randrup},
\begin{eqnarray}
\zeta_{f}(T)
	&=& \sum_i g^{\left(i\right)}_{f}
	\exp\left(-E^{*\left(i\right)}_{f}/T\right)
\nonumber\\
	&\simeq&
		g_f^{(g.s.)}
		+ {c_1 \over A_f^{5/3}}\ \int_0^\infty d E^* e^{-E^*/T}\  
			\exp(2\sqrt{a_f E^*})
\ ,\\
\nonumber
a_f	&=&
		{A_f \over 8}\,\left(1 - c_2 A_f^{-1/3}\right)
	\ (\hbox{MeV}^{-1})
\ ,\quad
c_1 = 0.2\ (\hbox{MeV}^{-1})
\ ,\quad
c_2 = 0.8
\ ,
\end{eqnarray}
where $g^{\left(i\right)}_{f} = 2j^{\left(i\right)}_{f} + 1$ 
is the spin degeneracy of the energy level at excitation energy
$E^{*(i)}_f$
of the fragment species $f$.

In NSE, the nuclear binding energy $B_f$ plays an essential role.
Fragment yields are sensitive to 
the binding energy modification due to, for example,
the medium effects in supernova matter. 
In studies of heavy-ion collisions, 
since the density and its fluctuation are large,
the repulsive interfragment Coulomb potentials
are taken into account explicitly
rather than in the form of mass modification.
In supernova matter under consideration,
attractive electron-fragment Coulomb potential effects are more important.
Since electron density is not negligible and almost constant,
we ignore inter-fragment Coulomb potentials
and the electron effects on intrinsic fragment Coulomb energies are 
incorporated in the form of binding energy modification.
%
We have used the Wigner-Seitz approximation 
in evaluating the Coulomb energy correction~\cite{Lattimer}
to the binding energy of nuclei adopted in NSE.
We assume that
the electrons are distributed uniformly
in a sphere with radius $R_{ef}$ which is determined to cancel
the charge of the fragment $f$ at a given electron density
$\rho_e$.
\begin{eqnarray}
B_f(\rho_e) &=&
	B_f(0) - \Delta V_f^{Coul}(\rho_e)
\ ,\\
\Delta V_f^{Coul} &=&
	-  \frac{3}{5}\frac{Z_f^2e^2}{R_0}
	\left(\frac32\,\eta_f - \frac12\,\eta_f^3\right)
\ ,\quad
    \eta_f \equiv \frac{R_{0f}}{R_{ef}} = 
	\left(\frac{\rho_e}{Z_f \rho_0/A_f}\right)^{1/3}
\ ,
\end{eqnarray}
where $B_f(0)$ is the nuclear binding energy in vacuum,
and $R_{0f}$ is the nuclear radius.

It is important to note that the Coulomb energy correction,
$\Delta V_f^{Coul}(\rho_e)$,
contains the term proportional to $\rho_e^{1/3}$.
Because of this functional form,
the correction is meaningfully large even at $\rhoB = 10^{-6}\rho_0$.
For example, when the electron fraction is $Y_e = 0.3$ ,
the reduction of the Coulomb energy 
for heavy nuclei amounts to 90 \% of the total Coulomb energy
at $\rhoB = \rho_0$,
and the reduction is around $10\ \MeV$ even at $10^{-6}\rho_0$.
The increase of binding energy acts to enhance heavy nuclei,
and some nuclei beyond the dripline at vacuum
or unstable against fission,
can be stabilized in supernova matter.
Finite gas nucleon density also plays a role to form nuclei
beyond the dripline tentatively
by the balance of nucleon absorption and emission.
We have adopted the mass table of Myers and Swiatecki~\cite{MS1994},
which is based on the Thomas-Fermi model with shell correction
for about 9000 kinds of nuclei.

Since nuclear binding energies depend on the electron density,
we have to solve the chemical equilibrium
condition of nuclei and leptons in supernova matter
in a consistent way 
to satisfy $F_\mu = \mu_p + \mu_e - \mu_n - \mu_\nu = 0$.
Provided that the baryon density and temperature are given,
and that the charge and lepton densities are fixed
as $(\rhoC, \rhoL) = (0, Y_L\rhoB)$, 
all the particle densities
are determined if the average proton fraction ($Y_p$) is given,
\begin{subeqnarray}
\label{eq:Yp-rhon}
(1 - Y_p)\, \rhoB
	&=& \sum_f N_f \rho_f(\mu_n, \mu_p, B_f(\rho_e))
	\equiv \overline{\rho}_n (\mu_n, \mu_p, \rho_e)
\ ,\\
\label{eq:Yp-rhop}
Y_p\, \rhoB
	&=& \sum_f Z_f \rho_f(\mu_n, \mu_p, B_f(\rho_e))
	\equiv \overline\rho_p (\mu_n, \mu_p, \rho_e)
\ ,\\
\label{eq:Yp-rhoeu}
Y_p\, \rhoB
	&=& \rho_e (\mu_e)
\ ,\quad
(Y_L - Y_p)\, \rhoB
	= \rho_{\nu} (\mu_\nu)
\ ,
\end{subeqnarray}
where $\overline\rho_n$ and $\overline\rho_p$ are the neutron and proton 
densities including those in nuclei.
We can easily solve last two conditions in 
Eqs.~(\ref{eq:Yp-rhoeu}) numerically,
and derivatives of $\mu_e$ and $\mu_\nu$ with respect to $Y_p$
can be obtained as
$d\mu_e/dY_p = \rhoB\,(d\rho_e/d\mu_e)^{-1}$
and 
$d\mu_\nu/dY_p = - \rhoB\,(d\rho_\nu/d\mu_\nu)^{-1}$.
Once $Y_p$ is given,
first two equations are the same
as usual conditions in fragment-based statistical models.
By using the charge neutrality condition $\rho_e = Y_p \rhoB$,
we can get the derivatives of $\mu_n$ and $\mu_p$
with respect to $Y_p$ as, 
\begin{equation}
\pmatrix{
	\begin{array}{cc}
	\partial \overline\rho_n / \partial \mu_n\ ,  &
	\partial \overline\rho_n / \partial \mu_p    \\
	\partial \overline\rho_p / \partial \mu_n\ ,  &
	\partial \overline\rho_p / \partial \mu_p    \\
	\end{array}
}
\pmatrix{
	\begin{array}{c}
	d\mu_n\\ 
	d\mu_p\\
	\end{array}
}
= \rhoB\ dY_p 
\pmatrix{
	\begin{array}{c}
	-1 - \partial \overline\rho_n / \partial \rho_e \\
	 1 - \partial \overline\rho_p / \partial \rho_e \\
	\end{array} 
}\ .
\end{equation}
Therefore, we can solve the chemical equilibrium condition, $F_\mu = 0$,
by the Newton's method, 
$\delta F_{\mu}= -F_{\mu}/\left(dF_{\mu}\left(Y_p\right)/dY_p\right)$.

\subsection{Gas Fraction and Equation of State in a Statistical Model}
\begin{figure}[tbhp]
\centerline{~\Psfig{6.0cm}{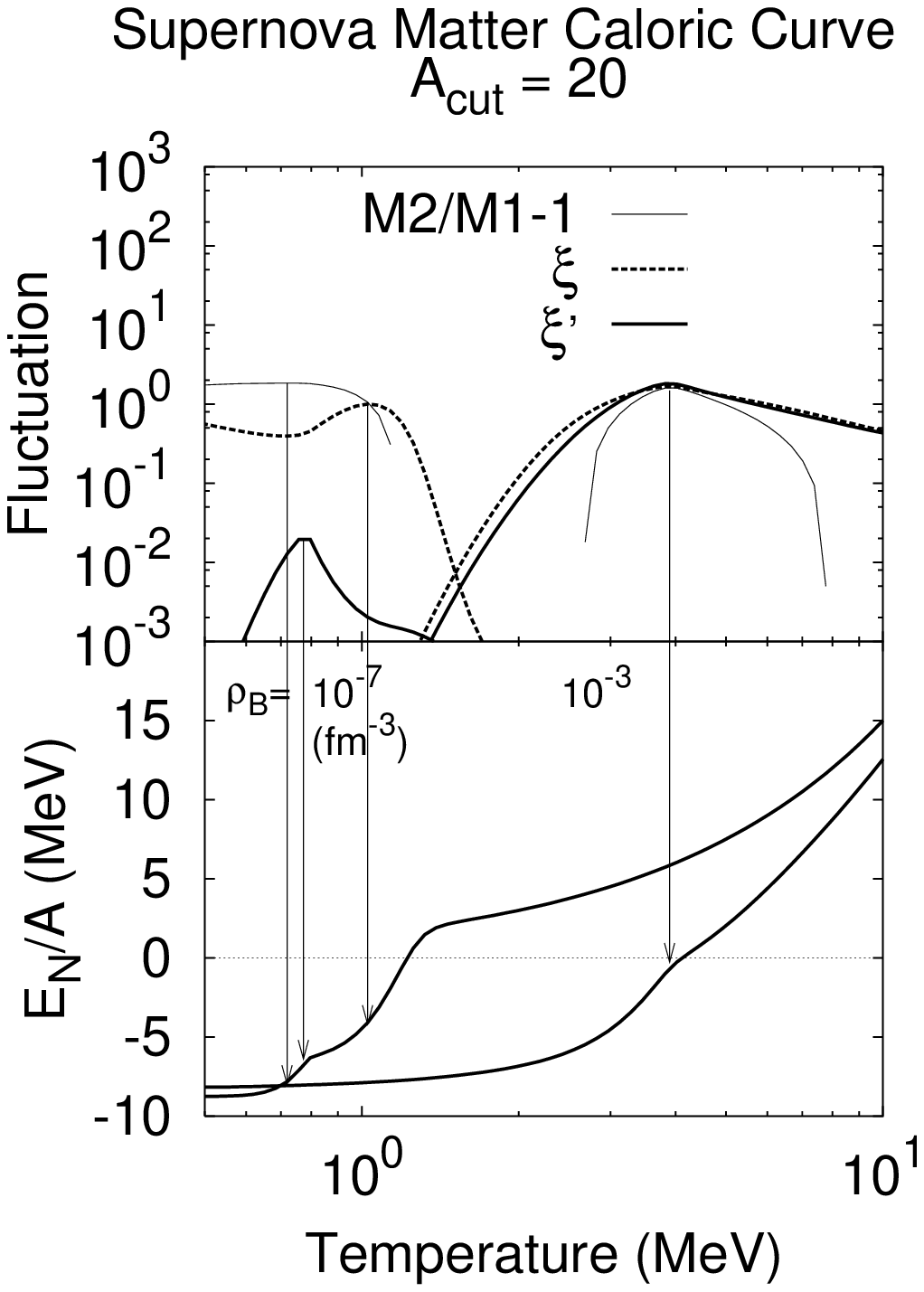}~\Psfig{6.0cm}{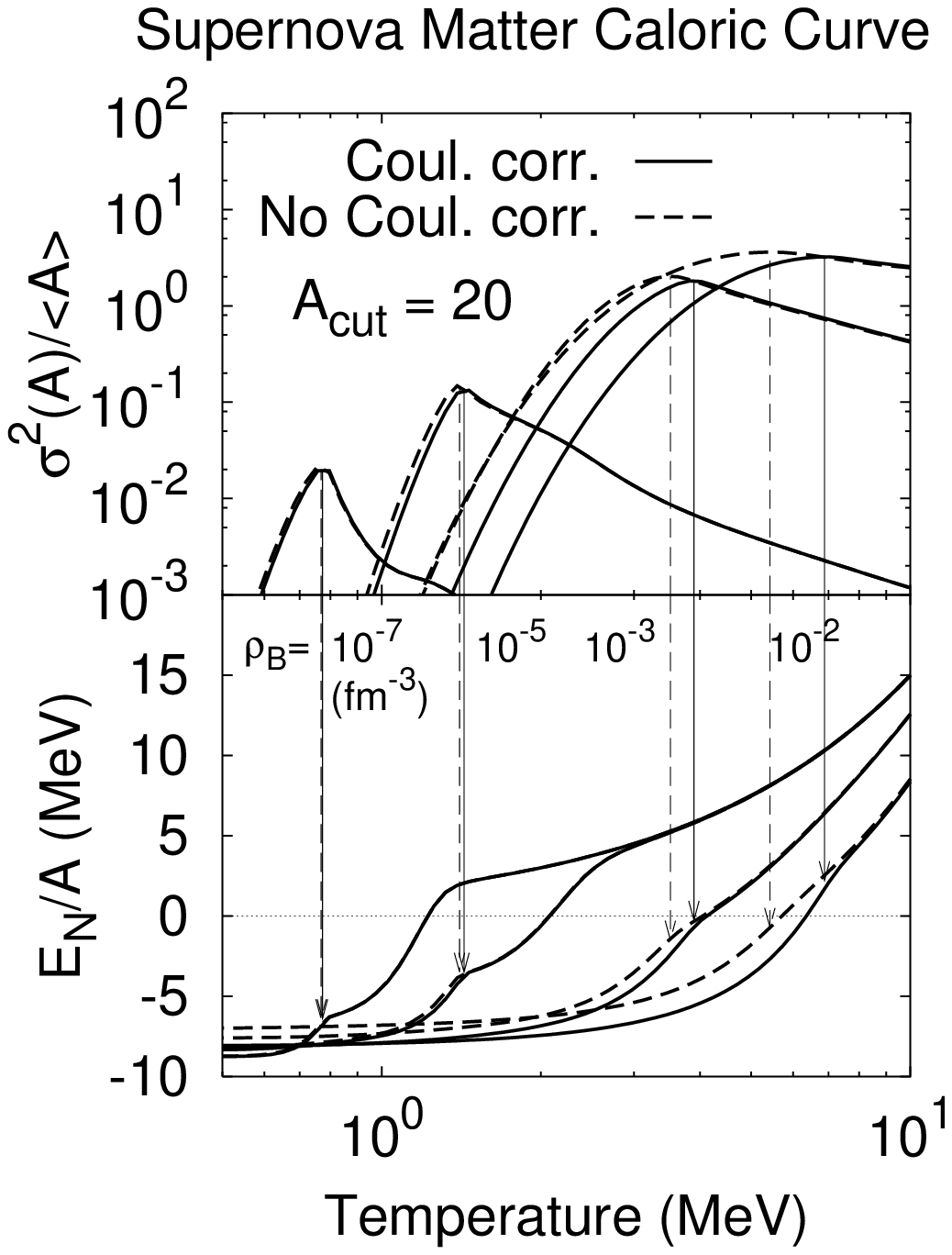}~}
\caption{
Left: Mass moment ratio (upper panel)
and caloric curve (lower)
of supernova matter with $Y_L = 0.35$ 
at $\rhoB = 10^{-7}$ and $10^{-3}\ \fmcube$.
In the upper panel, mass moment ratios, $M_2/M_1$ (thin solid lines),
$\xi = \sigma^2(A_f)/<A_f>$ (with $\alpha$, dotted lines),
$\xi'$ (same as $\xi$ but without $\alpha$, thick solid lines)
are shown.
Right: 
$\xi'$ (upper panel) and caloric curve (lower)
of supernova matter with $Y_L = 0.35$
at baryon densities $\rhoB = 10^{-7}, 10^{-5}, 10^{-3}$ and $10^{-2}\ \fmcube$.
Results with Coulomb correction (solid lines) and without Coulomb correction
(dashed lines) are compared.
}\label{Fig:CalMm}
\end{figure}
%
\begin{figure}[tbhp]
\centerline{~\Psfig{6.0cm}{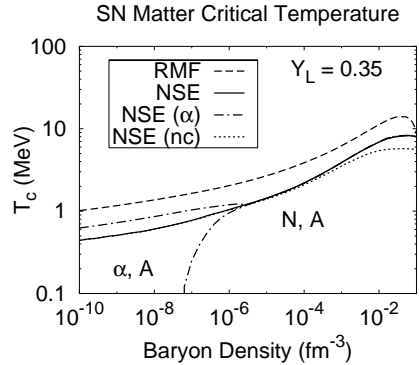}~}
\caption{
Liquid-gas coexisting region of supernova matter with $Y_L = 0.35$
calculated in NSE (solid line) in comparison with
RMF results (dashed line).
Results without Coulomb correction is shown by dotted curve.
Boiling point in the statistical model has been determined 
as the maximum point of mass variance-to-average ratio,
while there appear two local maxima in the fragment mass fluctuation
when we include the contribution of $\alpha$ particle,
as shown by dot-dashed curves.
} 
\label{Fig:Tc-NSE}
\end{figure}


%
%
Compared to RMF, there is no sharp phase transition in NSE
because of the finite size of fragments.
Although we can see kink-like behavior
in the nuclear part of energy per baryon as a function of the temperature
(caloric curve), this kink is not clear enough to define the boiling point
as seen in Fig.~\ref{Fig:CalMm} .
There are several definitions of $T_{boil}$ proposed in the literature.
For example,
$T_{boil}$ is proposed to be well defined at the peak of
$M_2/M_1$ by Bauer~\cite{Bauer}.
The $n$-th moment of light fragments, $M_n$,
is defined as $M_n \equiv \sum_f A_f^n\rho_f$.
Another way is to define $T_{boil}$ by the peak of
the light fragment mass variance-to-average ratio,
$\xi \equiv \sigma^2(A_f)/<A_f>$, 
which becomes small when one fragment (or nucleon) dominates,
and becomes unity when the mass distribution is a Poisonnian.
We show these ratios in the upper-left panel of Fig.~\ref{Fig:CalMm}.
At high densities, these definitions give reasonable boiling points,
but at low densities, the peak of $M_2/M_1$ becomes dull,
and $\xi$ shows two peaks.
From fragment distributions, we find that the two peak structure
appears due to the formation of $\alpha$ which can be comparable
to neutrons
at very low densities, where gas part becomes almost symmetric.
In Fig.~\ref{Fig:Tc-NSE}, we show the temperatures at local maxima
of $\xi$ by dot-dashed lines.
This formation of $\alpha$ makes also $M_2/M_1$ peak dull.
Therefore, we here define the boiling point 
as the peak of the light fragment mass variance-to-average ratio $\xi'$,
where we excluded $\alpha$ particle in the calculation of 
the light fragment mass average and variance.
As shown in the solid lines in Fig.~\ref{Fig:CalMm}, 
the peak position of $\xi'$ is well defined at any density.
In addition, this boiling point corresponds to the kink position 
in the caloric curve.

%
%
In Fig.~\ref{Fig:Tc-NSE}, 
we show the density dependence of $T_{boil}$, 
in comparison with the RMF results.
We find that the boiling points in NSE
are lower than those in the two phase
treatment of the RMF model by about a factor of two.
The reduction of $T_{boil}$ is a natural consequence of finite size of nuclei.
Because the intrinsic Coulomb energy cannot be completely removed by electrons
at densities $\rhoB < \rho_0$,
nuclei are limited to have finite size.
Then nuclei loses surface energy in addition to the Coulomb energy,
and gas nucleons are favored.
However, it is worthwhile to note that 
the qualitative behavior of $T_{boil}$ is similar to that in RMF,
and they are still high enough, $T_{boil} > 0.7$ MeV 
for $\rhoB > 10^{-7}\ \fmcube$.
We tabulate the boiling points at $\rhoB = 10^{-7}, 10^{-5}, 10^{-3}$
and $10^{-2}\ \fmcube$ in Table~\ref{table:Tc-NSE}.


%
\begin{table}
\caption{Boiling points in NSE
at $Y_L = 0.35$ as a function of the baryon density.
Results with (NSE) and without (NSE, nc) Coulomb corrections
are shown.
}\label{table:Tc-NSE}
\begin{center}
\begin{tabular}{r|cccc}
\hline
  $\rhoB (\fmcube)$ & $10^{-7}$ & $10^{-5}$ & $10^{-3}$ & $10^{-2}$ \\
\hline\hline
  $T_{boil}$(NSE) (MeV) &  0.77 &  1.43 &  3.90 &  6.88 \\
  $T_{boil}$(NSE,nc) (MeV) &  0.77 &  1.40 &  3.52 &  5.42 \\
\hline
\end{tabular}
\end{center}
\end{table}

\begin{figure}[tbhp]
\centerline{~\Psfig{8.0cm}{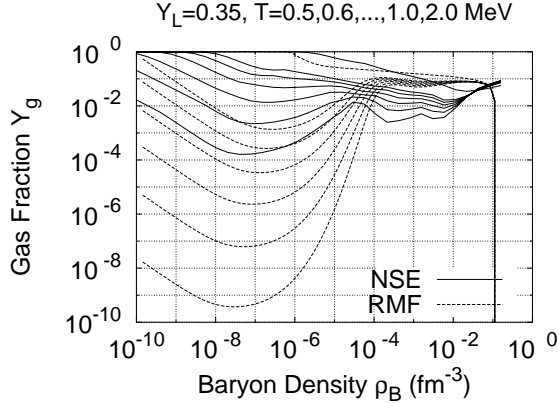}~}
\caption{
Gas fraction in supernova matter as a function of the baryon density
at $Y_L = 0.35$.
Results in NSE (RMF) are shown by solid (dashed) lines.
Temperatures are taken to be $T = 0.5, 0.6, ..., 1.0$ and $2.0$ MeV
from bottom to top.
}\label{Fig:Yg-NSE}
\end{figure}

%
%
In the RMF treatment, one of the most characteristic features in
the coexisting region is the reduction of the gas fraction.
This also applies to the NSE results.
We have defined the gas fraction 
as the ratio of isolated nucleon density to the total baryon density,
$Y_g \equiv (\rho_p + \rho_n) / \rhoB$.
As shown in Fig.~\ref{Fig:Yg-NSE},
gas fraction behaves similarly to that in RMF results;
as the baryon density decreases,
it quickly becomes very small until $\rhoB \sim 10^{-8}\ \fmcube$,
and gradually grows at lower densities again.
In addition, it is interesting to note that
$Y_g$ curves within the statistical model
have the second minimum at $\rhoB \sim 10^{-3}\ \fmcube$.
Supernova matter favors nuclear fragment
(nucleus) state rather than nucleon gas at these densities.
We call these minimum regions as 
the first ($\rhoB \sim 10^{-7}\ \fmcube$)
and the second ($\rhoB \sim 10^{-3}\ \fmcube$) fragment windows,
respectively.
The first one is caused by the drastic isospin-symmetrization of supernova
matter by leptons at low baryon densities.
The second fragment window is 
specific to NSE.
The mechanism of this appearance is not very clear,
but we find that
these two minima converges to one
when we use the liquid drop mass formula for the nuclear binding energies
and ignore the surface term.

\begin{figure}[tbhp]
\centerline{~\Psfig{8.0cm}{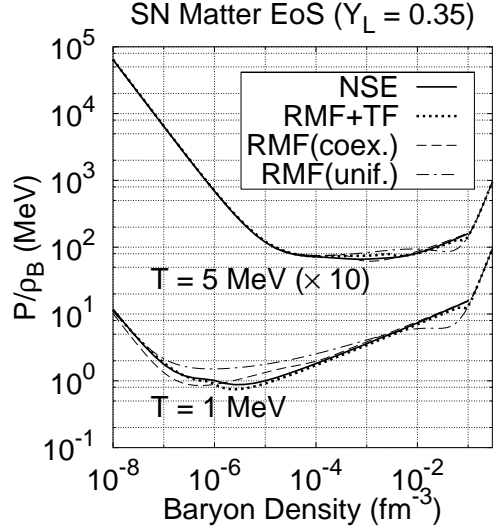}~}
\caption{
Equation of state of the supernova matter in NSE 
in comparison with those in RMF models
at $T=1$ MeV (lower curves)
and $T=5$ MeV (upper curves, scaled up by a factor of 10).
The solid, dotted, dashed and dot-dashed line denote
EoS in NSE,
RMF with Thomas-Fermi approximation model (RMF+TF)
\protect{\cite{TM1-table}},
the liquid-gas coexisting RMF model (RMF(Coex.)) and homogeneous 
RMF model (RMF(Unif.)),
respectively.
} 
\label{Fig:EoS-comp}
\end{figure}

%
%
In Fig.~\ref{Fig:EoS-comp},
we show the EoS in NSE at $T = 5$ and 1 MeV with $Y_L = 0.35$,
in comparison with those in 
RMF with the Thomas-Fermi approximation
for a dominant configuration~\cite{TM1-table} (RMF+TF, dotted lines),
RMF with two-phase coexistence (RMF(coex.), dashed lines),
and homogeneous RMF (RMF(unif.), dot-dashed lines).
We find that the finiteness of nuclei does not affect the EoS
at high densities, $\rhoB \geq 10^{-2}\ \fmcube\ (10^{-4}\ \fmcube)$
for $T = 5$ MeV (1 MeV),
but modifies the density dependence of the pressure at lower densities,
as seen in the difference between RMF(coex.) and RMF+TF.
Compared with the EoS in RMF+TF, 
the present NSE results give very similar pressures,
except for the density region $\rhoB \sim 10^{-3}\ \fmcube (10^{-6}\ \fmcube)$
for $T =$ 5 MeV (1 MeV), 
where the pressure are different by 10 $\sim$ 20 \%.
These densities correspond to the region where
the given temperatures are close to the boiling points.

The agreement of EoS in NSE and RMF+TF is somewhat surprising.
There are three large differences in NSE and RMF+TF:
(1) Nuclear masses are taken from the table~\cite{MS1994} in NSE,
while masses are calculated in RMF+TF.
(2) Interfragment and nucleon-fragment nuclear interactions 
are neglected in NSE, 
while they are included in RMF+TF.
(3) One configuration is assumed in RMF+TF,
while statistical ensemble is considered in NSE.
Thus the above agreement might suggest that
once nuclear masses are properly included,
nuclear interactions between gas nucleon and fragments
play a minor role in EoS.

\subsection{Fragment Distribution and Coulomb Correction Effects}

In this subsection,
we investigate fragment distribution in the density range
$10^{-7} \leq \rhoB \leq 10^{-2}\ \fmcube$,
starting from the first fragment window
to the density close to the critical point.
We are most interested in the boundary
of coexisting region
at the temperatures around $T_{boil} (\rhoB)$.
If the density is not very small,
the freeze-out temperature $T_{fo}$ is expected to be around or just
below the boiling point $T_{boil}$.
Below the boiling points, 
fragments are formed abundantly by absorbing many of gas nucleons,
and nuclear
{\em number} density of particles (sum of nucleon and fragment densities)
becomes quickly smaller,
then the mean free path for each particle becomes much longer.
This rapid fragment formation makes the typical
interaction intervals longer, and is expected to help 
the system to freeze-out.
The condition of freeze-out should be studied more carefully.

\begin{figure}[tbhp]
\Psfig{14cm}{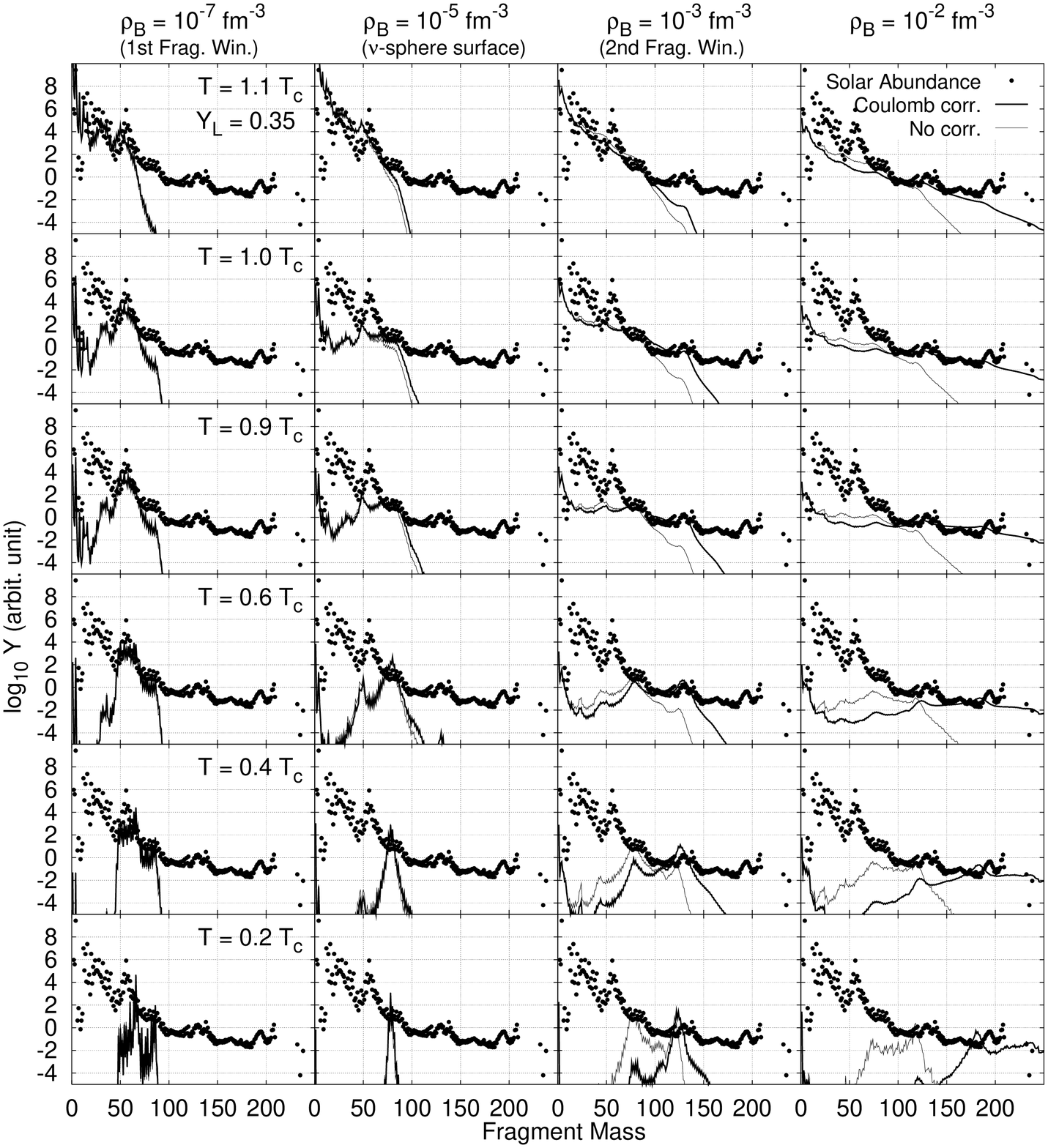}
\caption{
The fragment distribution at
$\rhoB= 10^{-7}\ \fmcube$ (first column), 
$10^{-5}\ \fmcube$ (second column),
$10^{-3}\ \fmcube$ (third column),
and 
$10^{-2}\ \fmcube$ (fourth column)
with lepton fraction $Y_L = 0.35$
in comparison with the observed solar abundance (dots).
Temperatures are taken to be $T=0.2\ T_{boil} \sim 1.1\ T_{boil}$,
where $T_{boil}$ is the boiling point at each density.
Thick and thin lines denote the distribution
with and without Coulomb correction, respectively.
}\label{Fig:SNA}
\end{figure}

We can see common features of the temperature dependence
of fragment mass distributions in each row of Fig.~\ref{Fig:SNA}.
When the temperature is higher than the boiling point,
the fragments almost obey the exponential distribution.
At the temperature near the boiling point,
the distribution shows the power-law like behavior up to some mass.
%
%
A similar trend, $Y(A_f) \sim A_f^{-\tau}$, also appears in nuclear 
multifragmentation~\cite{HI-Exp},
where $Y$ and $A_f$ denote the fragment yield and mass.
This power law was suggested by Fisher~\cite{Fisher}\ for a mass distribution
of droplets at around the critical point.
As the temperature becomes lower than the boiling point,
the distribution at each density becomes localized to some mass number.
The localization is caused by the small entropy and shell effects.

%
%
The density $\rhoB = 10^{-7}\ \fmcube$ corresponds to the
the first fragment window, where supernova matter is symmetrized by leptons
and nuclei are formed abundantly at low temperatures.
As shown in Fig.~\ref{Fig:SNA} (first column),
the main products near the boiling point
($T = T_{boil} \sim 0.77\ \MeV$)
are nucleon, $\alpha$, iron peak nuclei, and they extend to 
the first peak nuclei of r-process in the distribution shoulder.
The Coulomb correction effect is negligible at this density.
Although the above distribution seems to be similar
to the initial seed nuclear composition
in a standard scenario of the r-process,
we would like to point out that when the temperature
rises to around $T_{boil} \sim 0.8$ MeV,
the distribution become broad due to thermal fluctuation.
Especially, just above the boiling point, $T = 1.1\ T_{boil}$ (top panel), 
the equilibrium nuclear distribution resembles to the solar abundance
up to the iron peak.

%
%
Next, we show the results at $\rhoB = 10^{-5}\ \fmcube$
in Fig.~\ref{Fig:SNA} (second column),
which corresponds to the density around the neutrino sphere.
Most stable nuclei at this density
are those at around the first peak of r-process, 
as can be seen in the distribution at low temperatures.
As a result,
these nuclei are formed easily also at around the boiling point 
($T = 1.0\ T_{boil} = 1.43\ \MeV$),
in addition to nucleons, $\alpha$, and iron peak nuclei.
The Coulomb correction effect is not large,
but at around $T_{boil}$, we can see small enhancement of heavy fragments
with larger masses over the first peak of r-process.

%
%
In the second fragment window ($\rhoB = 10^{-3}\ \fmcube$),
power-law like behavior can be seen at around $T_{boil} = 3.90\ \MeV$
up to the second peak of r-process, 
as shown in the third column of Fig.~\ref{Fig:SNA}.
Coulomb correction effects are clearly seen at this density.
The center of the distribution shifts
from the first peak of r-process without Coulomb correction
to the second peak of r-process with Coulomb correction.

%
%
As shown in the fourth column of Fig.~\ref{Fig:SNA},
the distribution at $\rhoB = 10^{-2}\ \fmcube$ shows the same trend.
It is interesting to find that the peak position in the NSE results
is shifted downwards compared to the observed third peak of r-process.
This shift mainly comes from the $np$ ratio of formed nuclei.
The observed third peak of r-process is a consequence
of the neutron magic number $N = 126$
and the $np$ ratio along the r-process path.
In the present NSE model calculation,
having a large $np$ ratio,
nuclei beyond the neutron dripline
appear easily at equilibrium.

\begin{figure}[tbhp]
\centerline{~\Psfig{14.0cm}{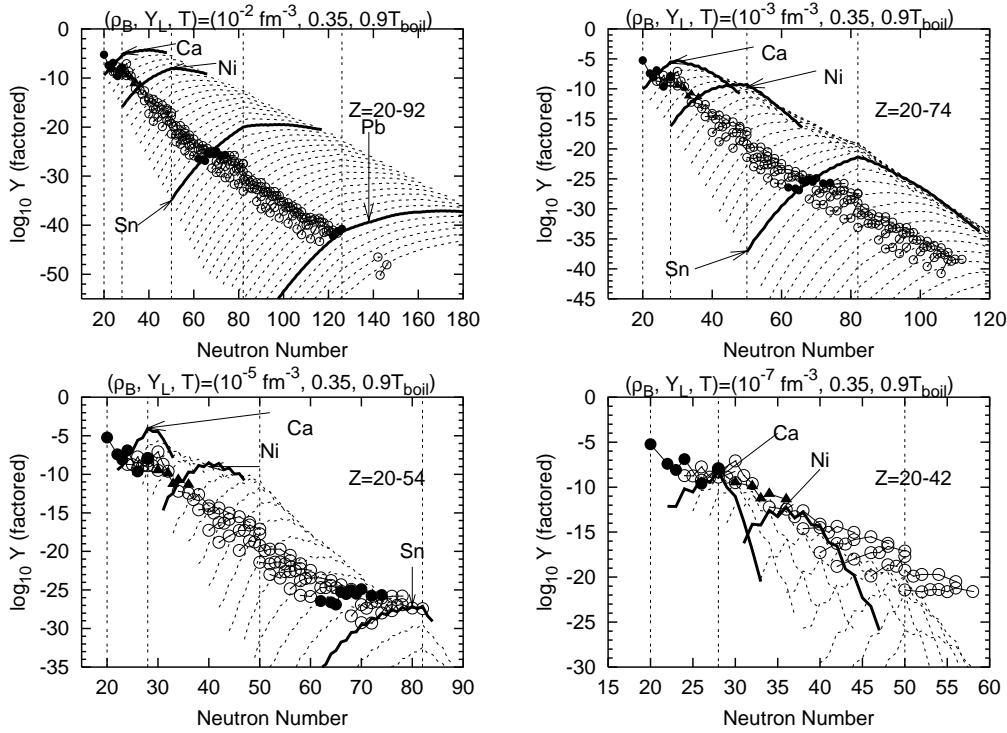}~}
\caption{
Calculated even $Z$ isotope distribution at
$\rhoB = 10^{-2}, 10^{-3}, 10^{-5}$ and $10^{-7}\ \fmcube$
and $(T, Y_L) = (0.9\ T_{boil}, 0.35)$
in comparison with the solar abundance by circle and
triangle symbols.
The boiling points are those for given densities.
Yields are shifted, for clarify of plot,
by a factor of 10 from $Z = 20$ (Ca)
to larger $Z$ nuclei (from top to bottom).
One overall factor is chosen to get good global fit.
}
\label{Fig:ZN}
\end{figure}

%
%
We also give an example of isotope distribution
at $Y_L = 0.35$ in Fig.~\ref{Fig:ZN}.
Here, we choose a slightly lower temperature than the boiling point,
$T_{fo} = 0.9\ T_{boil}$, for the freeze-out temperature,
as discussed at the beginning of this subsection.
The upper-left, upper-right, lower-left
and lower-right panels of Fig.~\ref{Fig:ZN} show
the isotope distributions at
$\rhoB = 10^{-2}, 10^{-3}, 10^{-5}$ and $10^{-7}\ \fmcube$, respectively.
At $\rhoB = 10^{-2}\ \fmcube$, 
while the observed isotope ratio are well explained in the calculation
from $Z = 20$ (Ca) to $Z= 92$ (U) with one overall normalization factor, 
much more neutron rich nuclei appear at equilibrium.
For isotones with $N = 126$, nuclei with $Z = 54$ to $Z = 92$ are formed.
This large $np$ ratio, which comes from the large $np$ asymmetry of the liquid
phase as shown in Fig.~\ref{Fig:Yp}, gives smaller mass number with $N = 126$.

At lower densities,
similar trends can be seen in the mass number range produced 
at around $T_{boil}$, 
except for $\rhoB = 10^{-7}\ \fmcube$.
As already discussed, nuclear matter is symmetrized outside of the neutrino
sphere, then the calculated distribution at $\rhoB = 10^{-7}\ \fmcube$
is shifted
toward proton rich side of the observed solar abundance
for large $Z$.

\subsection{Fragment Distribution outside of Neutrino Sphere
and in Neutrino-less Supernova Matter}

\begin{figure}
\centerline{\Psfig{14.0cm}{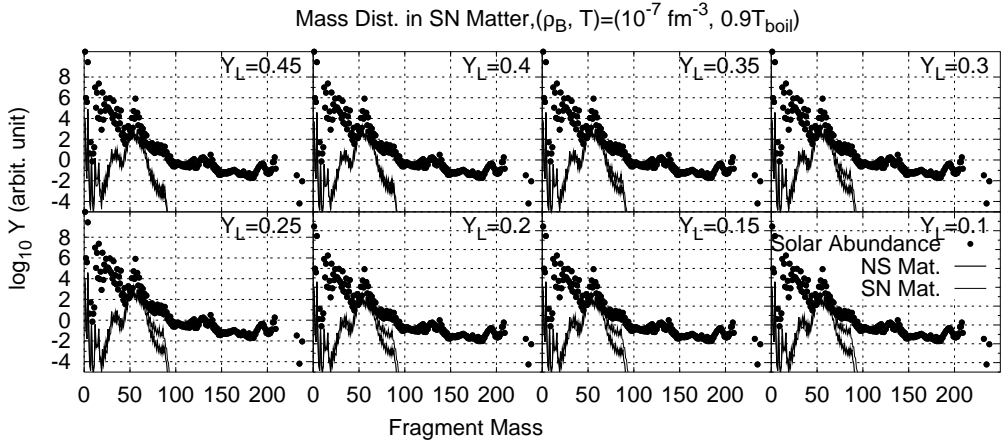}~}
\caption{
The fragment distribution
in neutrino-less supernova matter with $ Y_p = 0.46$ (thick lines, $n,p,e^-$)
and supernova matter (thin lines, $n,p,e^-,\nu_e$)
at $\rhoB = 10^{-7}\ \fmcube$
in comparison with the solar abundance (filled circles).
For supernova matter, we have chosen $Y_L = 0.1 \sim 0.45$.
The boiling points are almost independent on $Y_L$,
$T_{boil} \sim 0.77\ \MeV$.
} 
\label{Fig:SNL2_freeU}
\end{figure}

%
%
While we have considered the $\beta$-equilibrium condition with fixed
lepton fraction $Y_L$, proton fraction $Y_p$ is almost independent
on $Y_L$ outside of the neutrino sphere.
Once $Y_p$, $\rhoB$, and $T$ are given,
neutron and proton chemical potentials are uniquely determined,
thus boiling point and fragment distribution are also obtained uniquely.
In RMF, 
this independence on $Y_L$ outside of the neutrino sphere
can be found in $Y_p$ (Fig.~\ref{Fig:Yp})
and in $T_{boil}$ (Fig.~\ref{Fig:Tc}).
The same tendency
applies to fragment distribution
in the NSE results of supernova matter
and neutrino-less supernova (NS) matter
in which neutrino chemical potential is zero ($\mu_\nu = \muL = 0$).
In Fig.~\ref{Fig:SNL2_freeU}, 
we show the fragment mass distribution
in NS matter at $\rhoB = 10^{-7}\ \fmcube$,
in which $Y_p = 0.46$,
in comparison with that in supernova matter ($Y_L = 0.1 \sim 0.45$).
We find that the boiling points and nuclear distribution are almost
independent on $Y_L$,
and they are almost the same as those in NS matter.

The above observations tell us that the fragment distributions
under thermal and charge equilibrium
at around $T_{boil}$ are 
independent of degree of neutrino trapping.
In addition, the consequent fragment distribution 
provides iron-group nuclei,
which become seed elements in a standard r-process scenario.
On the contrary, $Y_p$ clearly depends on $Y_L$ inside the neutrino sphere.
The nuclear distribution formed in the neutrino sphere
is sensitive to the dynamics of supernova explosion, 
especially on how much neutrinos escape before the neutrino trapping.

\section{Relation to Nucleosynthesis in Supernovae}

One of the key questions on the LG-process 
is whether the elements made through the liquid-gas phase
transition are ejected to outside.
If they are ejected, they contribute to the following nucleosynthesis
such as the r-process.

As mentioned earlier,
all trajectories of adiabatic path go across the boundary of coexisting region.
If the material with $S/B \geq 10$ is ejected in supernova explosion, 
the present model calculation suggests that the material 
would go through the coexisting region before the ejection.  
This can happen if 
the freeze-out temperature of supernova matter is as low as
$T = (0.5-2)$ MeV
at densities $\rhoB$ = ($10^{-7} - 10^{-5}$)\ fm$^{-3}$,
where $S/B \geq 8$ at the boiling points
provided that the adiabatic paths go straight until the boundary.
Since the fragment distribution is almost independent of the lepton fraction 
at these densities, 
the distribution shown in Fig.~\ref{Fig:SNA} would appear regardless 
of details of neutrino trapping.  

In order to examine this possibility, we analyse the results of 
hydrodynamical calculation of core-collapse supernova~\cite{SumiyoshiNext}. 
In their calculation, the EoS table derived by the RMF-TF model 
with the TM1 interaction~\cite{TM1-table} has been used in 
the general relativistic hydrodynamics~\cite{Yam97}.
Note that this EoS table and the EoS in NSE give almost the same pressure
as shown in Fig.~\ref{Fig:EoS-comp}.
Among series of model calculations of adiabatic collapse,
a case of iron core of 15$M_{\odot}$ presupernova star~\cite{Woo95} 
explode hydrodynamically with the fixed electron fraction, 
which were assumed to study the properties of EoS in model explosion.  
In this model explosion, the material from deep inside the core 
can be ejected.
The entropy per baryon of this ejecta is about $S/B \sim 10$ 
determined by the shock passage.  
The trajectory of this ejecta passes through the coexisting region.
Further inner material has lower entropies down to $S/B \sim 1$, 
which are favorable for the LG process, but the ejection 
becomes rather difficult.  
It is also to be noted that the model explosion is obtained 
by a simplified calculation without neutrino-transfer 
assuming large electron fraction.  
Further careful studies are necessary to examine quantitatively 
the mass ejection by performing the numerical simulations 
of hydrodynamics with neutrino-transfer of core-collapse.

Although hydrodynamical explosion (so-called prompt explosion) 
has been claimed to be limited for very small iron core and extreme 
cases of EoS and electron capture rates~\cite{Bet90,Suz90}, 
the outcome of nucleosynthesis is extremely interesting 
if the mass ejection really occurs.  
A case of prompt explosion of the small iron core of 11$M_{\odot}$ 
presupernova star has been studied as a site of successful r-process 
nucleosynthesis by the dynamical ejection of neutron-rich 
material~\cite{Sumi01}.
In their studies, the material having entropy per baryon 
$S/B \sim 10$ is ejected.   
These ejecta may be affected by the LG-process to help the production 
of heavier r-process elements around actinides.  
If the ejection from relatively high density
$(\rhoB \geq 10^{-5}\ \fmcube)$ takes place, 
the formed fragments are generally very neutron rich and
some part of them are unstable against
neutron emission.
These nuclei provide a huge amount of neutrons, which help the following
r-process to proceed.
Mass ejection is also expected in asymmetric supernova explosion.  
The convection of material is generally seen in supernova simulations 
and the material deep inside may be ejected by hydrodynamical 
instability.  (See~\cite{Janka} for example.)  
Jet-like ejection of material may occur in rotating collapse of stars 
as discussed in some literatures~\cite{Shimizu,MacFadyen}.
The ejection of material and its consequence on nucleosynthesis 
should be clarified together with the problem of supernova explosion mechanism 
and require further extensive studies.  

\section{Summary and discussion}

In this paper, 
we have investigated the liquid-gas phase transition of supernova matter,
and its effects on the fragment formation.
We have used two models
---
the Relativistic Mean Field (RMF) model
and the Nuclear Statistical Equilibrium (NSE) model.

%
%
In RMF, we have used the interaction TM1, which has been successfully
applied to finite nuclei including neutron rich unstable nuclei,
neutron stars, and supernova explosion~\cite{TM1-table,Sum95b,Sum00,Sum95c}.
Leptons are shown to play non-trivial roles
such as the symmetrization of nuclear part of supernova matter.
As a result, nuclear liquid gains symmetry energy,
and the calculated boiling points in supernova matter 
($T_{boil} > 1\ \MeV$ for $\rhoB \geq 10^{-10}\ \fmcube$) are comparable
to those in symmetric nuclear matter at low densities.
Adiabatic paths are shown to go across the boundary of coexisting region 
even at high entropy such as $S/B \geq 10$, which is expected to be enough
for supernova matter to be ejected to outside.
Clear concentration of adiabatic paths to the boundary of coexisting region
have been found.
All of these findings suggest that at least a part of ejecta
in supernova explosion
would experience the liquid-gas phase transition before freeze-out.

%
%
In NSE, 
we have used nuclear binding energies of Myers-Swiatecki model~\cite{MS1994}
with Coulomb correction
due to electron screening as a medium effect~\cite{Lattimer}.
Since larger species of nuclei become stable with this Coulomb energy
correction, we have adopted the mass table
of around 9000 nuclei constructed by Myers and Swiatecki~\cite{MS1994}.
Because of the finiteness of nuclei,
they lose surface and Coulomb energy compared to the case of
coexistence treatment of two infinite matter phases in RMF.
The boiling points become slightly lower,
but they are still high; $T_{boil} \geq 0.7$ MeV for 
$\rhoB \geq 10^{-7}\ \fmcube$. 
Calculated fragment mass distributions around $T_{boil}(\rhoB)$
show enhancement of the iron peak elements,
the first, second, and third peak r-process elements
at $\rhoB = 10^{-7}, 10^{-5}, 10^{-3}$ and $10^{-2}\ \fmcube$,
respectively.
In addition, calculated isotope distribution shows that
very neutron rich nuclei around and beyond the neutron dripline may exist
under thermal and chemical equilibrium in supernova matter
with degenerate neutrinos.
These unstable nuclei against neutron emission
would provide a lot of neutrons after freeze-out,
which may help the r-process to proceed.
%

%
%
From the present investigations, we can draw a new scenario for
making seed nuclei before the r-process;
fragments are abundantly formed through the liquid-gas phase transition
of supernova matter before the freeze-out,
and this formation of fragments serve to produce the bulk structure
of the seed elements.
We call this process as the LG process
as a pre-process of r-process~\cite{IOS2001-YKIS01b}.

It is interesting to note that our model based on the liquid-gas
coexisting state of supernova matter can even provide
the r-process nuclei or their seed
in a simple manner based on the condition
determined by the dynamics of supernova explosion
such as $\rhoB$, $Y_L$, and $T$.
%
%
One of the most promising conditions is $\rhoB = 10^{-5}\ \fmcube$.
This density roughly corresponds to the neutrino sphere. 
The entropy at $T_{boil}$ 
is a little smaller than the ejection criteria, $S/B \geq 10$
in one-dimensional hydrodynamical calculation of supernova 
explosion~\cite{SumiyoshiNext}.
However, 
it would be possible that matter with small entropy can be 
ejected by convection and/or jet in asymmetric supernova 
explosion~\cite{Janka,MacFadyen}.
The most conservative freeze-out density for ejection
would be $\rhoB = 10^{-7}\ \fmcube$.
The entropy at $T_{boil}$ is large enough,
and the seed nuclei will be nucleons, $\alpha$, iron peak
nuclei and a small amount of the first peak nuclei of r-process.
Higher densities may not be relevant to ejection,
but it may be closely related to the nuclear distribution 
on hot neutron star surface.

%
%
In this work, we have assumed equilibrium throughout this paper.
One of the key questions is the freeze-out conditions of supernova matter,
at which nuclear reactions become less frequent
and supernova matter goes off equilibrium in the expansion time-scale.
The seed nuclear distribution of the r-process will be given as
the nuclear distribution on the freeze-out line in the $(\rhoB,T)$ 
diagram.
It is important to determine the freeze-out condition in supernova dynamics.
Another important direction is to construct a model
which includes both of the mean field nature such as in RMF
and the statistical nature in NSE. 
In a present NSE treatment, only the Coulomb correction is included
as the medium effects, and medium effects from strong interactions 
are neglected.
This neglection may lead to the overestimate of neutron rich nuclei,
as discussed in recent statistical fragmentation models~\cite{Stat-Isospin}.
On the other hand, in the Thomas-Fermi treatment of heavy-nuclei with EoS
derived using RMF, since
statistical nature or fragment distribution is not taken care of,
the treatment is not sufficient especially at around $T_{boil}$.
Works in these directions are in progress.


\section*{Acknowledgements}

The authors are grateful to
Prof. K. Kato, Dr. M. Ito and Dr. K. Iida
for useful discussions and suggestions.
This work was supported in part by
the Grant-in-Aid for Scientific Research
(Nos.\ 09640329
and 13740165) 
from the Ministry of Education, Science and Culture, Japan.

\newpage
\appendix

\section{Low Density Approximation}

In the calculation of liquid-gas coexistence at low temperatures,
it becomes necessary to solve the chemical equilibrium of liquid
and very low density gas.
For example, the gas baryon density
becomes around $\rhoB^{Gas} = 10^{-74}\ \fmcube$
at the coexisting condition $(\rhoB, T) =(10^{-10}\ \fmcube, 0.1 \hbox{MeV})$.
In order to efficiently obtain the derivative matrix
	$\partial \mu_i/\partial \rho_j$
($i,j = B, C, L$), which are required in solving coexistence, 
we take the low temperature and low density approximation
for nucleons at low baryon densities ($\rhoB \ll 10^{-5}\ \fmcube$)
around and below the boiling point 
in the mean field calculation.

In the mean field approximation, 
nucleon distribution is a function of
the effecitve mass
	$M^* = M + g_{\sigma} \sigma_0$
and effective chemical potentials
	$\nu_i = \mu_i - g_\omega \omega_0 - g_\rho \tau_i \rho_{30}$
	($i = n \hbox{ or } p$),
where $\sigma_0, \omega_0$ and $\rho_{30}$ are
the expectation values of the meson fields of 
$\sigma, \omega$ and the neutral $\rho$ mesons, respectively.
For a given baryon density $\rhoB$ and proton fraction $Y_p$,
the vector meson expectation values are uniquely deterimed as
$\omega_0 = \omega_0(\rhoB), \rho_{30} = \rho_{30}(\rho_\Ss{T})$,
where $\rho_\Ss{T} \equiv \rho_p - \rho_n$.
In a non-relativistic limit~\cite{Muller},
we can take energy as $E^* = M^* + p^2/2M^*$.
At low densities $\left(\rhoB \ll 10^{-5}\ \fmcube\right)$,
the nucleon fugacity $f_i \equiv \exp\left(-(M^*_i-\nu_i)/T\right)$ 
becomes much smaller than unity,
then we can safely ignore the second and higher order terms
in the fugacity $f_i$.
In this approximation, 
the Fermi distribution is approximated to be the Boltzmann distribution,
then 
	the baryon number density $\rho_i$
	and the scalar density $\rho_s$
are analytically obtained for a given value of $\sigma_0$ as,
\begin{eqnarray}
\label{Eq:AppRho}
\rho_i (\nu_i, \sigma_0)
	&=& G_-(\nu_i, M^*(\sigma_0), T) + O(f_i^2)
\ ,\\
\label{Eq:AppRhoS}
\rho_s (\nu_n, \nu_p, \sigma_0)
	&=&
		  G_+(\nu_n, M^*(\sigma_0), T)
		+ G_+(\nu_p, M^*(\sigma_0), T)
		+ O(f^2)
\ , \\
\label{Eq:AppG}
G_\pm (\nu, M^*, T)
	&=& 
	g \left(\frac{M^*T}{2 \pi \hbar^2}\right)^{\frac{3}{2}}
	\left(e^{\nu/T} \pm e^{-\nu/T}\right) e^{-M^*/T}
\ .
\end{eqnarray}
The self-consistent condition
$\sigma_0 = \sigma_0(\rho_s(\nu_n, \nu_p, \sigma_0))$
can be solved by interation using (\ref{Eq:AppRho}-\ref{Eq:AppG}),
which converges in a few steps at low densities.
All of the above densities are represented by three variables,
$\nu_n, \nu_p$ and $\sigma_0$ for a given temperature $T$,
and we can eliminate the $\sigma_0$ dependence
by using the total derivative of the above self-consistent condition.
\begin{eqnarray}
\label{Eq:AppDSigma}
d\sigma_0 &=& \frac{d\sigma_0}{d\rho_s}
	\left(
	  \frac{\partial\rho_s}{\partial\nu_n}\,d\nu_n
	+ \frac{\partial\rho_s}{\partial\nu_p}\,d\nu_p
	+ \frac{\partial\rho_s}{\partial\sigma_0}\,d\sigma_0
	\right)
\ ,\\
\label{Eq:AppDRho}
d\rho_i &=& \frac{\partial\rho_i}{\partial\nu_i}\,d\nu_i
	+ \frac{\partial\rho_p}{\partial\sigma_0} d\sigma_0
\ ,\\
\label{Eq:AppDMu}
d\mu_i &=& d\nu_i
	+ g_\omega \frac{d\omega_0}{d\rhoB}\,d\rhoB
	+ g_\rho \tau_i \frac{d\rho_{30}}{d\rho_\Ss{T}}\,d\rho_\Ss{T}
\ .
\end{eqnarray}
By solving the first equation (\ref{Eq:AppDSigma}) in $d\sigma_0$
and substituting it in the second equation (\ref{Eq:AppDRho}),
we obtain $\partial\rho_i/\partial\nu_j$
and then $\partial\nu_i/\partial\rho_j$.
Finally, the third equation (\ref{Eq:AppDMu}) gives 
the partial derivatives $\partial\mu_i/\partial\rho_j$
by eliminating $d\nu$.
Having
the (relativistic) lepton integrals 
and the above derivatives in nucleons,
it is straightforward to construct
the matrix $\partial\mu_i/\partial\rho_j$ ($i,j = B, C, L$).

\section{Procedures to obtain coexisting region of supernova matter in RMF}

The boundary of coexisting region of supernova matter has been determined
in three steps; symmetric nuclear matter, asymmetric nuclear matter,
and supernova matter.
%

\begin{figure}[bhtp]
\centerline{~\Psfig{6.0cm}{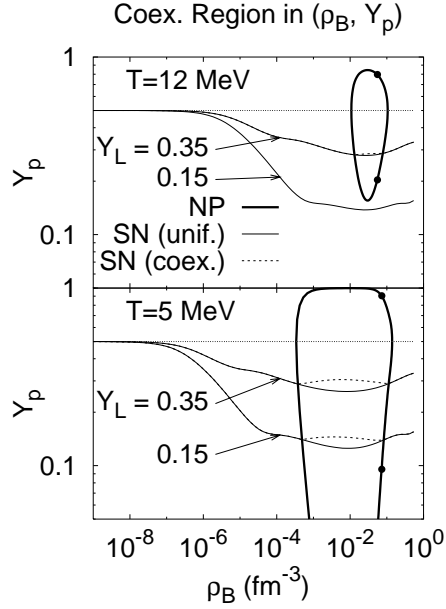}~}
\caption{
The boundary of
liquid-gas coexisting region in nuclear matter (thick solid lines),
and proton fraction as a function of the baryon density in supernova matter
without (thin solid lines) and with (dotted line) coexistence.
Filled circles show the points where liquid and gas phases converges to
the uniform matter.
}\label{Fig:NPcxYL}
\end{figure}

%
First, for a given temperature,
we solve the coexisting condition in symmetric nuclear matter
($Y_p^{Liq.}=Y_p^{Gas}=0.5$).
If there is a density region where pressure is decreasing
for increasing $\rhoB$,
liquid and gas phases can coexist,
and we can find coexisting densities, $\rhoB^{Liq.}$ and $\rhoB^{Gas}$, 
by using the Maxwell construction.
%
%
Secondly, we solve the coexisting condition for
	$(\rhoB^{Liq.}, Y_p^{Liq.})$
	and
	$(\rhoB^{Gas}, Y_p^{Gas})$
in asymmetric nuclear matter.
Having the coexisting condition at a given $Y_p^{Liq.}$,
it is easy to find coexisting condition for slightly different $Y_p^{Liq.}$
by using the multi-dimensional Newton's method.
Starting from symmetric nuclear matter,
three variables ($\rhoB^{Liq.}, \rhoB^{Gas}, Y_p^{Gas}$) are determined
for a given $Y_p^{Liq.}$ which is slightly different
from that in the previously solved condition
by requiring the condition,
	$\muB^{Liq.} = \muB^{Gas}$,
	$\muC^{Liq.} = \muC^{Gas}$,
	and 
	$P^{Liq.} = P^{Gas}$.
We show the boundary of coexisting region              of $(\rhoB, Y_p)$
by the thick solid line in Fig.~\ref{Fig:NPcxYL}.
Filled circles show the point where two phase become uniform,
$(\rhoB^{Liq.}, Y_p^{Liq.})=(\rhoB^{Gas}, Y_p^{Gas})$.
The density gap $\rhoB^{Liq.}-\rhoB^{Gas}$ generally decreases
as the liquid becomes more asymmetric, because of the symmetry energy loss.
At lower temperature, the coexisting pressure and thus 
the coexisting gas baryon density become small,
then larger density gap appears.
%
%
Thirdly,
proton fraction $Y_p$ is determined as a function of $\rhoB$ in 
uniform (homogeneous) supernova matter at a given $Y_L$,
by using the charge neutrality condition,
	$\rho_e = \rho_p = Y_p \rhoB$,
and the chemical equilibrium condition,
	$\mu_\nu = \muL = \mu_e + \mu_p - \mu_n$.
When $Y_p$ is in the coexisting region of nuclear matter,
liquid and gas phases can coexist in supernova matter.
The boundary of coexisting region               
for a given $T$ is determined 
by the crossing point of these two lines.
Since $Y_p$ becomes smaller for smaller $Y_L$ as shown in Fig.~\ref{Fig:NPcxYL},
the coexisting density region becomes narrower for smaller $Y_L$.
This is the reason why 
the boiling temperatures decrease for smaller lepton fraction
as shown in Fig.~\ref{Fig:Tc}.

For neutrino-less supernova matter (NS),
the procedure is almost the same,
except that the chemical equilibrium condition is modified 
to $\mu_\nu = 0$.

\end{document}